# Episodic Star Formation in the Carina dSph Galaxy[1]

T. A. Smecker-Hane

*University of California, Dept. of Physics & Astronomy, Irvine, CA, 92717, USA*

P. B. Stetson, J. E. Hesser

*Dominion Astrophysical Observatory, 5071 W. Saanich Rd., RR 5, Victoria, BC V8X 4M6, Canada*

D. A. VandenBerg

*University of Victoria, Dept. of Physics & Astronomy, P.O.B. 3055, Victoria, BC V8W 3P6, Canada*

**Abstract.** $B$, $R$ photometry for stars in the Carina dwarf spheroidal galaxy (dSph) that is unprecedented in depth, accuracy and spatial coverage demonstrates that it has had a surprisingly complex evolution dominated by episodic bursts of star formation. Comparing the color-magnitude diagram (CMD) with new theoretical isochrones, we determine that the ages of the "bursts" are approximately 2, 3–6, and 11–13 Gyr. Note that "bursts" and quiescent phases lasted for $\gtrsim 1$ Gyr, which are much longer than the $\sim 0.1$ Gyr dynamical timescale. Each main-sequence turnoff (MSTO) in Carina connects to a single red giant branch, which tells us that, regardless of age, Carina stars are metal-poor with [Fe/H]$\simeq -1.86$ and a spread in metal abundance of $\sigma_{[Fe/H]} \leq 0.2$ dex. Thus star formation in, and the loss of metal-enriched gas from, Carina proceeded slowly over many Gyr.

## 1. Introduction

The dwarf galaxies in the Local Group offer unique opportunities for testing the importance of dark matter halos and galactic winds in shaping the evolution of dwarf galaxies. Because of their proximity, we can make star-by-star measurements that disentangle the degeneracy of age and metallicity. Undoubtedly, supernovae have powered large-scale galactic winds which expelled the interstellar medium (ISM) and halted star formation in today's gas-poor dwarf spheroidals (dSphs) and dwarf ellipticals (dEs) (Dekel & Silk 1986, Vader 1986, Silk, Wyse & Shields 1987). However, we still do not know the timescales on which this

---

occurs, nor do we fully understand the physical processes which determine it. In the last decade, we began to appreciate that even some of the low mass dSphs have had complex star-formation rates (SFRs). This began with the discovery of carbon stars in numerous dSphs and the Andromeda II dE (*e.g.,* Aaronson & Mould 1985, Aaronson *et al.* 1985) and the observation of sizeable internal dispersions in the chemical abundances of dSphs (see reviews by Suntzeff 1988 and Da Costa 1988). Color-magnitude diagrams (CMDs) that reached the main-sequence turnoffs of dSphs began to reveal how complex their SFRs have been. For example, the CMD and luminosity function of the Carina dSph revealed that its SFR was dominated by two episodes of star formation roughly 7.5 and 15 Gyr ago (Mighell 1990, Mighell & Butcher 1992). Although CMDs alone cannot unravel a complex SFR because isochrones are degenerate in distance, age, metallicity and reddening, spectroscopic chemical abundances in concert with CMDs can. Therefore, with the current generation of large format, high-quantum efficiency CCDs and multi-fiber spectrographs on large ground-based telescopes, as well as with the Hubble Space Telescope, we now have the tools to determine the evolution of the nearby dwarf galaxies.

## 2. The Carina dSph: Observations and CMD

We used the CTIO 1.5m telescope to perform a wide-field photometric survey of $\sim 90\%$ of all the giants in Carina (Smecker-Hane, *et al.* 1994). We discovered two, morphologically-distinct, horizontal branches (HBs). An old HB and RR Lyrae instability strip belongs to a $\gtrsim 10$ Gyr stellar population, and a populous red clump HB belongs to an intermediate-age population. Using HB stars as tracers of the two populations, we found no difference in their radial distributions and inferred that star formation in the two bursts occurred over similar spatial extents. A new distance modulus of $(m-M)_0 = 20.12 \pm 0.04$ was derived from the magnitudes of the old HB and the tip of the red giant branch.

In addition, we obtained photometry of the MSTO region of Carina with the CTIO 4m telescope. The most stunning features in the CMD (see Figure 1) are the distinct, young MSTOs and the prominent gap in the SFR from approximately 6 to 11 Gyr ago. The median internal photometric errors at $R = 24$ are only $\sigma_R = 0.026$ and $\sigma_{B-R} = 0.043$. We are in the process of calibrating external errors, due, for example, to crowding, with extensive artificial star tests. From past experience, we anticipate that they will be $\lesssim 2$ times the internal errors, and we should be able to resolve the duration of the first burst of SF if it is $> 1$ Gyr.

In both studies, we achieved high photometry accuracy through careful data acquisition and new data reduction techniques. For the MSTO photometry, we used multiple exposures (20 $B$ images, 38 $R$ images) of a single field with moderate exposure times to build up an average of 5 hrs total integration time in each filter. Multiple observations yield robust measures of internal accuracies. Offsetting the telescope between exposures moves stars onto different parts of the CCD and thus decreases the effect of random flat field errors. We also obtain a dramatic decrease in photometric scatter using ALLFRAME, a new photometry algorithm (Stetson 1994). Positional information from *all frames* simulatneously is used to compute a single set of coordinates (as well as frame-to-frame coordinate transformations) and magnitudes for *each* star on *each* frame.

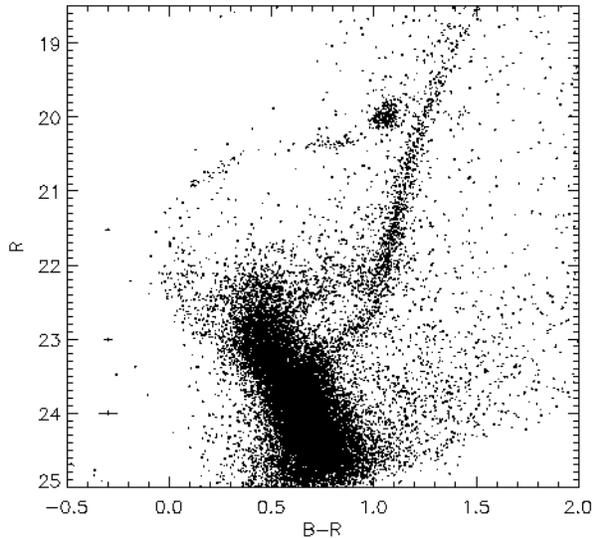

Figure 1.  The CMD of the Carina dSph. The median internal photometric error at magnitudes $R = 21.5, 23$, and $24$ are shown as on the left-hand side.

## 3.  Star Formation History

Figure 2 shows new theoretical isochrones from VandenBerg *et al.* (private communications, 1996). The multiple MSTOs in Carina have ages of approximately 2, 3–6 and 11–13 Gyr. Carina apparently depleted its gas slowly over many Gyr with long, few Gyr, gaps of inactivity. We will be comparing the CMD and luminosity functions with models to explicitly determine the SFR.

The isochrones were computed from new stellar evolutionary tracks of VandenBerg, Swenson, Rogers, Iglesias & Alexander. These tracks are computed with: OPAL opacities with a realistic elemental mix ($[\alpha/\text{Fe}] = +0.3$ for $[\text{Fe}/\text{H}] = -1.84$) as opposed to scaled-solar abundances, a non-ideal gas equation of state using the Debye-Hückel approximation, and up-to-date nuclear reaction rates. The new color-temperature relationship is based on model atmospheres with empirical corrections (when necessary) determined from globular cluster CMDs.

## 4.  Chemical Enrichment

All MSTOs in Carina connect to a single, narrow, red giant branch. Comparing with the theoretical isochrones, we estimate the mean metallicity is $[\text{Fe}/\text{H}] = -1.86$ with an internal dispersion of $\lesssim 0.2$ dex. This is similar to what Da

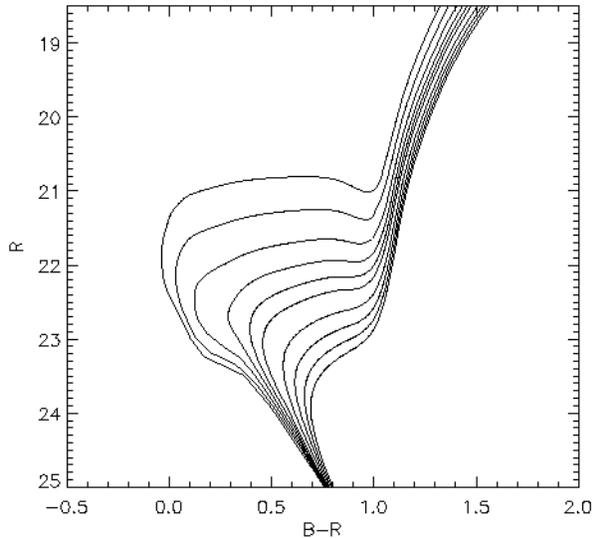

Figure 2. New theoretical isochrones from VandenBerg *et al.* (1996) for [Fe/H]= −1.84 and ages of 1.3, 2, 3, 4, 5, 6, 8, 10, 12, and 14 Gyr at the distance and reddening of the Carina dSph.

Costa and Hatzidimitriou have found from spectroscopy of the Ca II infrared triplet in 15 Carina stars (Da Costa 1994). However, they also found one star with [Fe/H]= −2.5 and more data are needed to determine the distribution of metallicities (and hence accurate ages) for the older, less numerous, population in Carina. We are collaborating with them to use the AAT 2dF spectrograph to obtain spectra for ∼ 150 giants in Carina.

## 5. Discussion

Carina has had little chemical enrichment although it has had numerous generations of stars with each episode lasting ≳ 1 Gyr (many times longer the dynamical timescale of ∼ 0.1 Gyr). This relatively gentle SFR continually generated winds which expelled metal-enriched supernovae ejecta. However, the wind apparently did not couple strongly to the ISM and the densest gas clouds were not easily expelled from the galaxy[2]. Thus the process of depleting the galaxy of its ISM occurred slowly over many Gyr. Massive dark matter halos inferred from the measured velocity dispersions (*e.g.,* Mateo *et al.* 1993) have been thought to provide stability and allow dwarf galaxies to remain intact even after copious gas loss. However if the gas is lost adiabatically, as in Carina, then dwarfs might not require a dark matter halo to be stable.

---

[2]There are problems with alternate scenarios. For example, additional multiple epochs of star formation might be triggered by the accretions of fresh gas. However, this would only be plausible if the dark matter halo of Carina were more massive than current estimates (1.1 × $10^7$ $M_\odot$; Mateo *et al.* 1993) and/or a population of cool gas clouds shared similar orbits around the Galactic halo at radii ∼ 100 kpc. Another scenario, in which gas flows out of Carina, cools as a result of work done against hot Galactic halo gas, and sinks back into the center, would result in metal enrichment that is *not* observed.

Carina is not unique in having multiple "bursts" of star formation. Leo I has distinct ages $\simeq 3$ and $\gtrsim 12$ Gyr with a stunning 90% of the stars being young (Mateo *et al.* 1995), and Sagittarus has ages $\simeq 4$ and $\gtrsim 10$ Gyr. The discrete ages of star clusters as well as field stars in the Large Magellanic Cloud suggest that it underwent surges in its SFR (see Girardi *et al.* 1995, Westerlund *et al.* 1995, and references therein). Nearby dwarf irregulars have had SFRS in the past $\lesssim 1$ Gyr which are either constant or nearly constant with with brief episodes of quiescence ("gasping" SFRs; Marconi *et al.* 1995, Tosi *et al.* 1991). The evidence points to a tight coupling between cooling and star formation and the disruptive feedback of supernovae explosions (at least in the absence of an external trigger such as a tidal disruption). The dIrs and dSphs appear to form a continuum in which the dSphs have less mass, a lower mean SFR, and more efficient gas loss. Hence dSphs simply run out of steam before the dIrs do.